\documentclass[onecolumn,amsmath,amssymb,aps,prd,showpacs,showkeys,nofootinbib]{revtex4}

\DeclareMathOperator{\Arctanh}{Arctanh}

\begin{document}

%\large

\author{V. S. Popov}
\author{M. A. Trusov}
\email{trusov@itep.ru}

\affiliation{ITEP, Moscow, Russia}

\title{Generating functions and sum rules for quantum oscillator}

\date{March 23, 2009}



\begin{abstract}
Generating functions and sum rules are discussed for transition
probabilities between quantum oscillator eigenstates with
time-dependent parameters.
\end{abstract}

\maketitle

The approach supposed by Feynman \cite{1} for disentangling of
non-commuting operators (FDM) was successfully applied to various
time-dependent problems in quantum mechanics \cite{2}--\cite{5}.
Using FDM one can \cite{1,3} calculate the transition
probabilities between eigenstates $|m,t\to -\infty\rangle$ and
$|n,t\to +\infty\rangle$ of a quantum oscillator with definite
initial and final quantum numbers, and compute corresponding
generating functions. If the oscillator has a constant frequency
$\omega$ being under an external force $f(t)$\footnote{A natural
condition $f(t)\to 0$ at $t\to\pm\infty$ is supposed.}, for the
generating function one has\footnote{This expression can be easily
derived \cite{5} from Schwinger result \cite{8} for a sum
$\sum_{m=0}^\infty w_{mn} u^m $.}:
\begin{equation}
G(u, v\mid \nu) = \sum^{\infty}_{m,n=0} w_{mn} (\nu) u^m v^n =
(1-u v)^{-1}\exp \left\{ -\nu~ \frac{(1-u)(1-v)}{1-uv}
\right\},\label{1}
\end{equation}
where $u$, $v$ are the auxiliary complex numbers and $\nu$ is the
oscillator excitation parameter:
\begin{equation}
\nu =\frac{1}{2\omega}\biggl\vert \int\limits^{\infty}_{-\infty}
f(t) \exp(-i \omega t) dt\biggr\vert^2,~~~\hbar=m =1.\label{2}
\end{equation}

%The problem was solved for the first time by Feynman using

From (\ref{1}) one obtains some non-trivial relations for
probabilities $w_{mn}$. First, integrating by $\nu$, one can
derive several peculiar sum rules
\begin{equation}
\begin{gathered}
\int\limits^{\infty}_0 w_{mn} (\nu) d\nu =1,\quad
\langle \nu \rangle_{mn}=\int\limits^{\infty}_0 w_{mn} (\nu)\nu d\nu = m+n+1,\\
\langle \Delta \nu^2 \rangle_{mn} =\int\limits^{\infty}_0 w_{mn}
(\nu) \left[\nu - \langle \nu \rangle_{mn}\right]^2 d\nu = 2mn +
m+n+1
\end{gathered}
\label{3}
\end{equation}
for arbitrary quantum numbers $m$ and $n$. For the initial vacuum
state ($m=0$) we have $\langle \Delta \nu^2 \rangle_{0n} = \langle
\nu \rangle_{0n}$ which corresponds to the Poisson distribution.

Taking in (\ref{1}) $u=v$ one has
\begin{equation*}
G(u,u|\nu)=(1-u^2)^{-1}\exp\left\{ 2\nu u(1+u)^{-1} -\nu \right\}
\end{equation*}
which yields
\begin{equation}
S_k (\nu) \equiv \sum_{m+n=k} w_{mn} (\nu) = e^{-\nu}
p_k(\nu),\label{4}
\end{equation}
where $p_k (\nu) =\sum_{s=0}^k (-1)^s L_s(2\nu)$ and $L_s$ are the
Laguerre polynomials, so
\begin{equation*}
p_0 =1,~~ p_1 = 2\nu,~~ p_2=2\nu^2-2 \nu +1,~~ p_3
=\frac{4}{3}\nu^3 - 4\nu^2 + 4\nu,~ \dots
\end{equation*}

For the oscillator with variable frequency $\omega(t)$ without an
external force\footnote{For this case the transitions occur only
between states with same parity.} ($f(t)\equiv 0$) one obtains
\begin{equation}
G(u,v\mid\rho) = \sum^{\infty}_{m,n=0} w_{mn}(\rho) u^m v^n
=\sqrt{(1-\rho)/\left[(1-uv)^2-\rho(u-v)^2\right]}, \label{5}
\end{equation}
and
\begin{equation}
\int\limits^1_0 G(u,v\mid \rho) (1-\rho)^{-1} d\rho = \frac{2}{u
-v}(\Arctanh u - \Arctanh v).\label{6}
\end{equation}
Here $\rho$, $0\le \rho \le 1$ is the oscillator excitation
parameter; details can be found in \cite{3,5}.  The frequency
$\omega(t)$ is an arbitrary real time function.  As usual, we
propose the boundary conditions:
\begin{equation*}
\omega(t)\to\omega_{\pm} \quad \text{at} \quad t\to\pm\infty
\end{equation*}
which allows one to define the final and initial eigenstates of
the oscillator. Note that the expressions (\ref{1}) and (\ref{5})
result from the general Husimi expression \cite{9}.

As a result we have
\begin{equation}
\begin{gathered}
\int\limits^1_0 \frac{w_{mn}(\rho)}{1-\rho} d\rho
=\frac{1+(-1)^{m+n}}{m+n+1},\\
\int\limits^1_0 \frac{w_{mn}(\rho)}{\rho\sqrt{1-\rho}} d\rho
=\frac{1+(-1)^{m+n}}{\mid m - n\mid},~~~m\neq n.
\end{gathered}
\label{7}
\end{equation}

Analogously to (\ref{3}) let us calculate the integral
$J_{mn}=\int\limits^1_0 w_{mn}(\rho) d\rho$. For diagonal ($m=n$)
transitions
\begin{equation}
J_{nn} =\frac{1}{2n+1} \biggl [ 1+ \frac{1}{(2n+3)(2n-1)}\biggr
],~~~n=0,1,2,..., \label{8}
\end{equation}
at $m\ne n$ the expression for $J_{mn}$ is  more cumbersome.

Finally, taking in (\ref{5}) $u=v$, one obtains
\begin{equation*}
G(u,u\mid\rho)=\frac{\sqrt{1-\rho}}{1-u^2}
\end{equation*}
and
%\begin{equation}
%S_k(\rho) =\sum_{m+n=k} w_{mn} (\rho) = \left\{ \begin{array}{ll}
%\sqrt{1-\rho}, &~~~k=0,2,4, ...\\ & \\ 0, &~~~k=1,3,5,
%...\end{array} \right.\label{9}
%\end{equation}
\begin{equation}
S_k(\rho) =\sum_{m+n=k} w_{mn} (\rho) = \begin{cases}
\sqrt{1-\rho}, & k=0,2,4,\dots \\ & \\ 0, & k=1,3,5, \dots
\end{cases} \label{9}
\end{equation}
(compare to (\ref{4})).

On differentiating subsequently the generating functions on
parameters $u$ and $v$, one can calculate the average quantum
number in the final state
\begin{equation}
\frac{\partial G}{\partial v} \biggl\vert_{v=1} =
\sum^{\infty}_{m=0} \langle n \rangle_m u^m,~~~\langle n \rangle_m
=\sum^{\infty}_{n=0} n w_{m
n}=-\frac{1}{2}+\left(m+\frac{1}{2}\right)\frac{1+\rho}{1-\rho},\label{10}
\end{equation}
the dispersion $\langle \Delta n^2 \rangle_m$ and other higher
distribution momenta.

In \cite{4} a more general model of a singular oscillator with
variable frequency was considered:
\begin{equation}
\begin{gathered}
\Hat H=\frac{1}{2}p^2+\frac{1}{2}\omega(t)^2x^2+\frac{g}{8x^2},\\
0<x<+\infty, \quad g=\text{const},\quad g>-1.
\end{gathered}
\label{a1}
\end{equation}

It is well known that at a fixed $t$ the instantaneous spectrum of
the Hamiltonian (\ref{a1}) is equidistant (see, e.g., \cite{11}):
\begin{equation}
E_n=2\omega(n-j),\qquad
j=-\frac{1}{2}-\frac{1}{4}\sqrt{1+g},\qquad n=0,1,2,\dots .
\label{a2}
\end{equation}

Using FDM, one can disentangle the operators in $H$ and, after
application of some group theory methods, calculate the transition
amplitudes between initial $|m\rangle$ and final $|n\rangle$
states, being expressed in terms of the generalized Wigner
function for the irreducible representation of the $su(1,1)$
algebra with weight $j$. As a result, one obtains explicit
analytical expressions for the transition probabilities $w_{mn}$.
The corresponding generating function takes the form:
\begin{equation}
g(u,v)=\sum_{m,n} w_{mn} u^m
v^n=\frac{\lambda^{-2j}}{1-uv\cdot\lambda^2} \label{13}
\end{equation}
where
\begin{equation*}
\lambda=\frac{2(1-\rho)}{1-\rho(u+v)+uv+\sqrt{\left[1-\rho(u+v)+uv\right]^2-4uv(1-\rho)^2}}
%\label{13p}
\end{equation*}
(see \cite{Trusov} for details).

One can show that at $j=-3/4$, which corresponds to the regular
oscillator ($g=0$) with odd levels, these expressions turn into
(\ref{5}).

%\newpage

The most interesting physical case is the excitation of the
oscillator ground level, i.e. $m=0$. Taking in (\ref{13}) $u=0$
and, correspondingly, $\lambda=(1-\rho)/(1-\rho v)$, one obtains
\begin{equation}
g(0,v)=\sum_{n=0}^{\infty} w_{0n} v^n=\left(\frac{1-\rho}{1-\rho
v}\right)^{-2j} \label{14}
\end{equation}
and
\begin{equation}
w_{0n}(\rho)=\frac{\Gamma(n-2j)}{n!\,\Gamma(-2j)}\rho^n(1-\rho)^{-2j},\quad
n=0,1,2,\dots \label{15}
\end{equation}

This formula also yields \cite{3} the transition probabilities of
the regular oscillator for $j=-1/4$ (even states, $0^{\text{th}}$
level excitation) and for $j=-3/4$ (odd states, $1^{\text{th}}$
level excitation).

To conclude, let us note that the generating function (\ref{13})
determines also the adiabatic expansion of the probability
$w_{mn}$ for the case of small values of the excitation parameter.
For example, for diagonal transitions ($m=n$) one has
\begin{equation}
w_{nn}(\rho)=1-2[n^2-(2n+1)j]\rho+\dots=1-\frac{1}{2}\left(N^2+N+1\right)\rho+\mathcal{O}(\rho^2)
\label{17}
\end{equation}
where
\begin{equation*}
N=2\sqrt{(n-j)^2-(j(j+1)+3/16)}-\frac{1}{2}.
\end{equation*}
In particular, $N=2n$ for $j=-1/4$ and $N=2n+1$ for $j=-3/4$.

\begin{acknowledgments}
This work was partially supported by the Russian Foundation for
Basic Research (grant No. 07-02-01116) and by the Ministry of
Science and Education of the Russian Federation (grant No. RNP
2.1.1. 1972). One of the authors (M.A.T.) also thanks for partial
support the President Grant No. NSh-4961.2008.2 and the President
Grant No. MK-2130.2008.2 .
\end{acknowledgments}

\end{document}